\documentclass[a4paper,11pt]{article}
\usepackage{pos}
\usepackage{wrapfig}

\input ArtNouvc.fd
\newcommand*\initfamily{\usefont{U}{ArtNouvc}{xl}{n}}
\newenvironment{test}%
  {\initfamily}%
  {}

\title{eXtreme19: when \begin{test} art\end{test} and \texttt{science} make the front page}
\ShortTitle{Extreme19: when art and science make the front page}

\author*[a]{Elisa Prandini}
\author[a]{Michele Doro}
\author[a]{Manuela Mallamaci}
\author[b]{Barbara Montolli}
\author[b]{Daria Mauri}

\affiliation[a]{Dipartimento di Fisica e Astronomia 'G. Galilei', Padova University,\\
  Via Marzolo 8, 35131, Padova, Italy}
\affiliation[b]{Liceo  Artistico Statale Amedeo Modigliani,\\ Via Enrico degli Scrovegni, 30, 35131 Padova, Italy}

\emailAdd{elisa.prandini@unipd.it}

\abstract{Back in mid 2018, we were organizing \textit{eXtreme19}, a conference on astro-particle physics held in Padova on the topic of extremely energetic emission from galaxies. For the preparation of the graphical material in support of the conference we seeked for a collaboration with talented \begin{test} art\end{test} students. To this purpose, we joined the Italian programme ‘PCTO’ (Percorsi per le Competenze Trasversali e per l’Orientamento) of high school student stages in job centers. Emily, Beatrice, and Chiara from the high school “Liceo Artistico Statale Amedeo Modigliani” in Padova accepted our invitation and started a 6-month stage at the Padova University in close contact with us. The challenge was to interbreed our \texttt{scientific} description of a relativistic jet of a powerful galaxy and their artistic assimilation and subsequent representation of it. During this period, they elaborated outstanding and innovative graphical material used for the webpage as well as the conference poster. The quality of the graphics was indeed excellent: one of their drawings became the cover of the February 2020 issue of the prestigious {\it Nature Astronomy} journal.}

\FullConference{37$^{\rm{th}}$ International Cosmic Ray Conference (ICRC 2021)\\
		July 12th -- 23rd, 2021\\
		Online -- Berlin, Germany}

\begin{document}

\maketitle

\section{Introduction}
The duty of preparation of local or international events such as workshops and conferences is part of the (almost) daily life of a \texttt{scientific} researcher. Results cannot be only shared through publications, but must be discussed in \texttt{scientific} gatherings. The success of the discussion at these meetings is not only in the quality of the speaker and the topics (although this constitutes the by-far main part) but also in the overall organization: clear travel information, clear schedule, coziness of places. One small but important point is also the publicity material, especially in early times, in which the attention of a researcher must be grasped (over often hundreds of mails and invitations). Nice conference logo and poster can therefore contribute to this task. Regardless the importance, a nice image is always a nice image and gives pleasure to the viewer. 

When we started organizing the international scientific conference \textit{eXtreme19} back in 2018, we were then wondering about our logo and poster image. The conference itself had the aim to gather experts on a niche topic, that of the so-called extreme blazars, a subclass of Active Galactic Nuclei (AGNs) displaying ultra-relativistic jets emerging from their central Super Massive Black Holes (SMBHs), and whose higher energy emission is peaked in the gamma-ray band. This class of objects is dubbed extreme in comparison with similar objects with different spectral energy distributions called low-peaked, medium-peaked, and high-peaked blazars. Figure~\ref{fig:blazar} shows an artistic sketch of a blazar and the  spectral energy distribution of the different blazar classes in which the 'Extreme' class is shown. 

Sometimes for a \texttt{scientist} it is difficult to envision \begin{test}art\end{test} images because of few factors. On one side clearly on average \texttt{scientist}s have limited skills in producing digital or analogical images. On the other hand, \texttt{scientist}s do love schematic views of \texttt{scientific} facts, and they are completely satisfied with that. Such feeling is clearly not shared by the average population. Any image produced by a \texttt{scientist} could suffer from this biased view. For such reasons, back then we discussed the opportunity to contact local \begin{test} artists\end{test} to design the poster image. We decided to make use of the Italian Ministry of Education's PCTO program (Percorsi per le Competenze Trasversali e per l’Orientamento). Such program obliges high school students of every cycle to devote a certain amount of their teaching time (of order of 1-2 hundreds of hours in total) to experiences outside school. This could encompass stages in industries, or at the Universities, among hundreds of possibilities. We therefore contacted the Liceo Artistico Statale (public High School of \begin{test} arts\end{test}) "Amedeo Modigliani”, located at walking distance from our Institute, and we got positive feedback from Prof. Montolli. She proposed our project to her students and she got positive feedback from three of them: Beatrice, Chiara, and Emily. We will not write their surnames here for privacy reasons. They were underage at that time, as they were attending the 3rd year of high school (that lasts 5 years in total). We started with meetings in person together also with Prof. Mauri of the same school, to discuss the project. The main idea was to develop a logo and a front image for the conference poster. About this, Emily says:

\begin{quote}
\textit{<<Initially I was worried about the idea of working on a subject unknown to me, such as astrophysics. However, I had the perception that it would be an important and enriching training experience. At a didactic level, this was a group project, and as such we would have needed to find a balance between creativity and taste. At a personal level, I had to confront with real commitments out of the school duties.>> [Emily] 
}
\end{quote}

Emily centered the topic. We did not want just to have an image, but to experience a project work with them. It would have been easier to assign, say, a logo to one person and an image to the other, but we ourselves wanted to learn from the process discussions.

\section{The poster and logo project}
\begin{figure}[h!t]
    \centering
    \includegraphics[height=4.5cm]{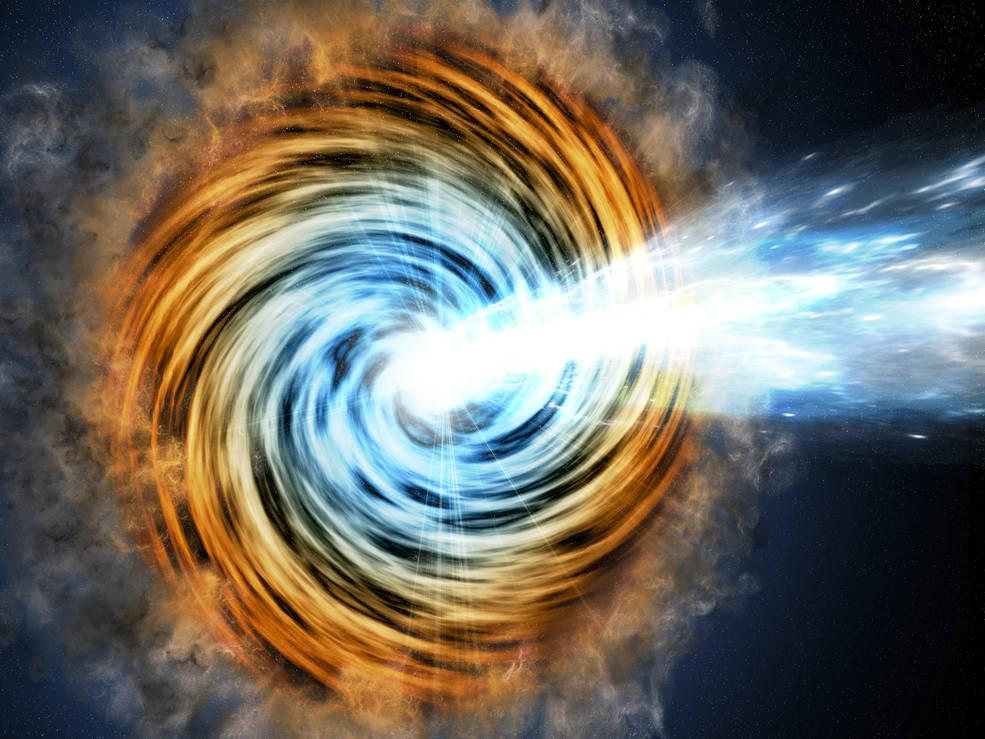}
        \includegraphics[height=4.5cm]{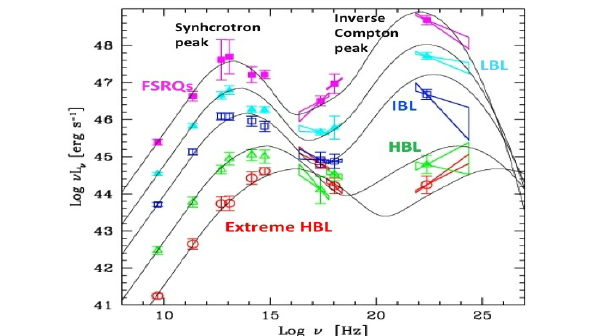}  
    \caption{(left) Wikipedia's image of a 'blazar'. (right) Spectral energy distributions of the different blazar classes. Image Courtesy: Falomo et al. (2014), Fossati et al. (1998).}
    \label{fig:blazar}
    \end{figure}

As mentioned above, the topic of the conference was a class of astrophysical targets, called extreme blazars. Images of blazars can be easily found by googling the word online, see e.g. Fig.~\ref{fig:blazar}. From the figure, one can grasp the basics of a blazar: a large amount of matter whirling around a SMBH (not visible in the image) and a powerful emission emerging from the central zone, perpendicular to the whirling plane, that is the blazar's ultra-relativistic jet. 
For the conference material, we did not want just to make yet another blazar picture, as there are many online. We wanted to focus on the topic of the conference that was on the multi-wavelength and multi-messenger observation of such targets. By multi-wavelength here we mean the observation of the same object in different energy bands (radio, UV, optical, X-rays, and gamma-rays). The knowledge about this class of targets can be obtained only if we study it with at all those wavelengths, and this requires the use of many different instruments, from radio telescopes to gamma-ray telescopes. Our targets can therefore be known only by putting together groups of experts from sometimes faraway instruments. Multi-messenger is a modern term explaining that the Universe is not only seen in its electromagnetic radiation but also through other messengers such as neutrinos and very recently from gravitational waves. Our students had therefore to express this 'multiplicity' of points of view in the image. Emily said:

\begin{quote}
\textit{<<The most complex part was the ideation of images that were both aesthetically interesting and scientifically accurate. We needed strong interaction and help from the University's experts to reach a final design.>> [Emily] 
}
\end{quote}

\begin{figure}[t]
\centering
\vspace{-10pt}
\includegraphics[width=0.9\linewidth]{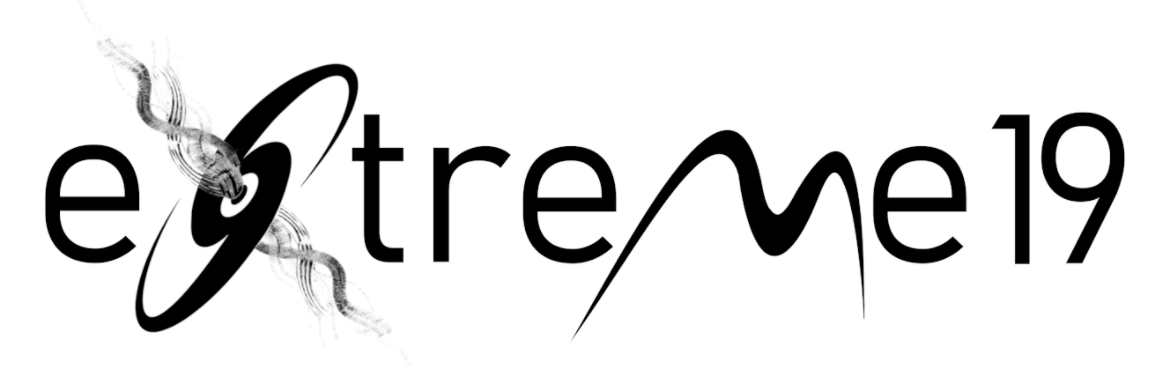} 
\includegraphics[width=0.9\linewidth]{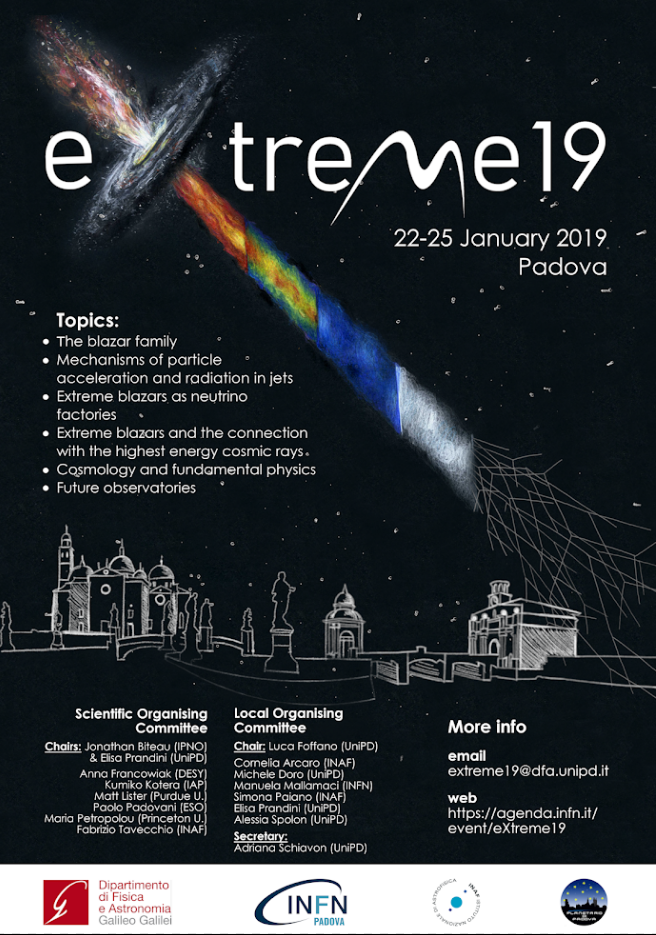} 
\caption{Top: The event logo for the \textit{eXtreme19} conference. The 'm'-shaped logo representing the blazar spectral energy distribution and the 'x' is shaped as the blazar itself. Bottom: \textit{eXtreme19} conference poster. One can see the main contributions: the ultra-relativistic blazar jet with both multi-wavelength and multi-messenger ideas, the logo and the skyline of Padova.}
\label{fig:poster}
\end{figure}

As one can see in Fig.~\ref{fig:poster}, the image produced by Beatrice, Chiara, and Emily mixed very well the \texttt{scientific} information with the pictorial \begin{test} art\end{test}. One one side of the blazar is the truer jet, with mixed particles and radiation, and on the other one can see the 'multi-wavelength and multi-messenger' nature displayed: the four coloured stripes represent four distinct wavelength ranges, and the last stripe represents the other messengers. 
\begin{test} hurray!\end{test}

Another important point of a poster is normally the communication of the location of the conference. We therefore discussed with the students how to communicate Padova and they came with the proposal of designing a sort of skyline with some basic Padova elements such as the notorius Sant'Antonio church, a monument in Prato della Valle, and the Porta Ognissanti.
The students also took care of adding the classic conference topic information such as topics, the \texttt{scientific} organizing committee, local organizing committee, and contacts, as well as the logos of the funding agencies.

Next, we wanted to have a conference 'logo', that is something to put on the conference webpage and on gadgets, or just to represent our conference at a glance. Nobody of us had previous expertise in logos. The students and their professors again found a perfect interbreeding of \begin{test} art\end{test} and \texttt{science}. The logo created by the students is displayed in Fig.~\ref{fig:poster}: the letter 'M" of the word 'extreme' in the conference is shaped as the blazar spectral energy distribution (see Fig.~\ref{fig:blazar}, right). Everybody in the field recognizes that 'M' as a double hump representing a blazar! And the letter 'X' of the word 'extreme' is shaped like a blazar. \begin{test} HURRAY!\end{test}

The conference \textit{eXtreme19}\footnote{The curious reader could find further information about the conference here: \url{https://agenda.infn.it/event/15975/}} was a success on all aspects, \texttt{scientific}, organizational and recreative. The participation was ample, with 70 \texttt{scientist} participants. 
The \texttt{scientific} success of the conference was certainly given by the quality of the speakers, but the overall success of the conference was surely determined by these other factors we were mentioning above. 

\section{Cover selection by the Nature Astronomy journal}
\begin{wrapfigure}{l}{0.65\linewidth}
\centering
\vspace{-10pt}
\includegraphics[width=0.95\linewidth]{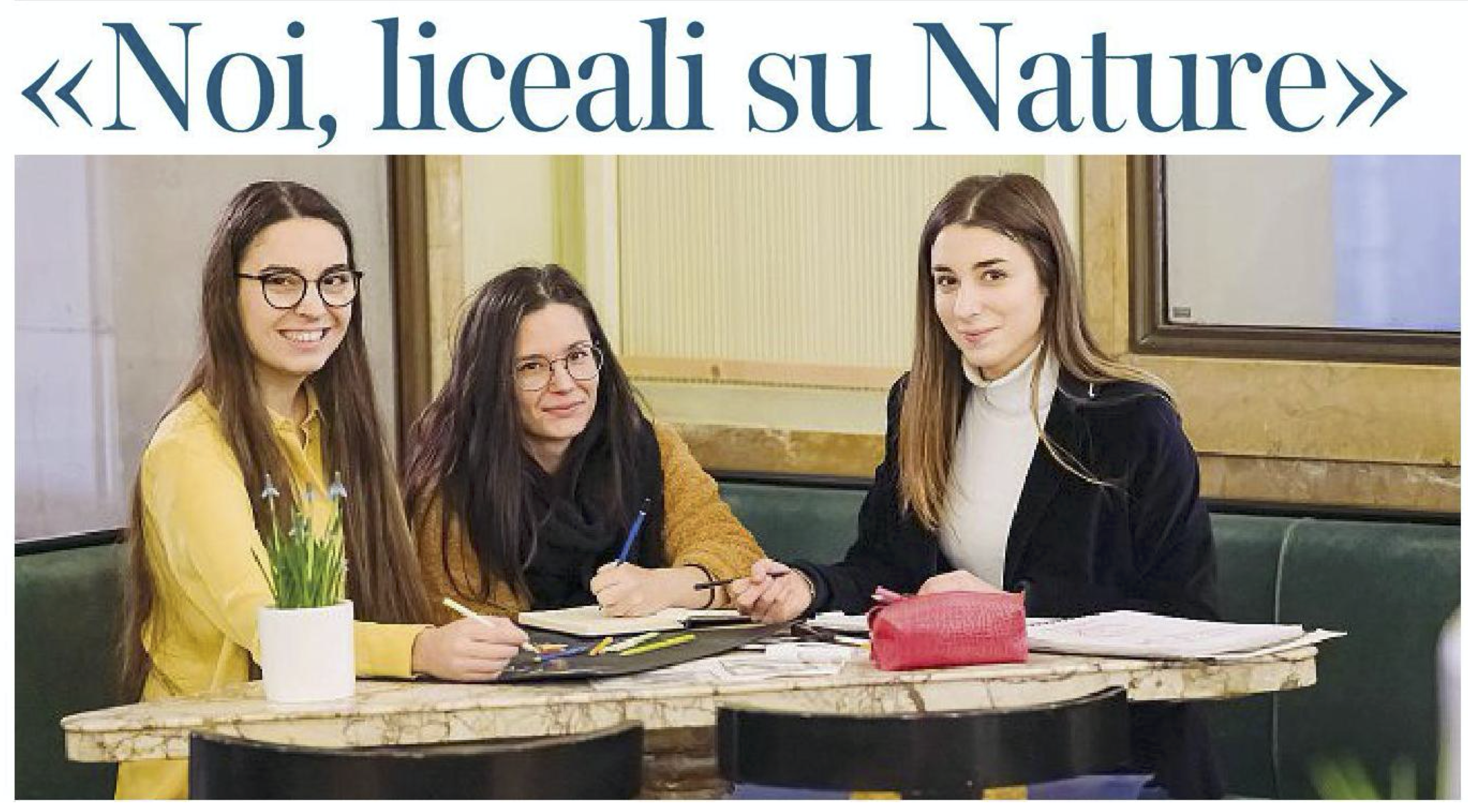}  
    \caption{Emily, Beatrice, Chiara appearing in the national newspaper \texttt{Corriere della Sera}.}
    \vspace{-10pt}
    \label{fig:corriere}
\end{wrapfigure}  
During the last day of the conference, Dr Prandini and Dr Biteau, co-organizer of the \texttt{scientific} part of \textit{eXtreme19},agreed that the discussions and results emerged at the conference were extremely interesting. Maybe a review paper, i.e. a paper that describes the different aspects and open questions of a \texttt{scientific} topic, could be proposed to a high-impact journal? 
With this idea in mind, they proposed to all invited speakers to participate in the review. Once the team of 8 researchers was formed, Biteau made the proposal to the prestigious journal Nature Astronomy and it was soon accepted as a perspective article! 

In late 2020, the perspective paper draft was accepted by the journal editor, who asked us if we had a proposal for the cover of the journal related to our work. We immediately thought that the material produced for \textit{eXtreme19} by Chiara, Emily, and Beatrice could be proposed, in particular the drawing used for the conference poster. It was a great honour for us all, \texttt{scientist}s, teachers, and students, when we discovered that our proposal was accepted and that we were going to make the front page of one of  the most prestigious journals in the astronomy and astrophysics field!

\begin{figure}
\centering
\includegraphics[width=0.95\linewidth]{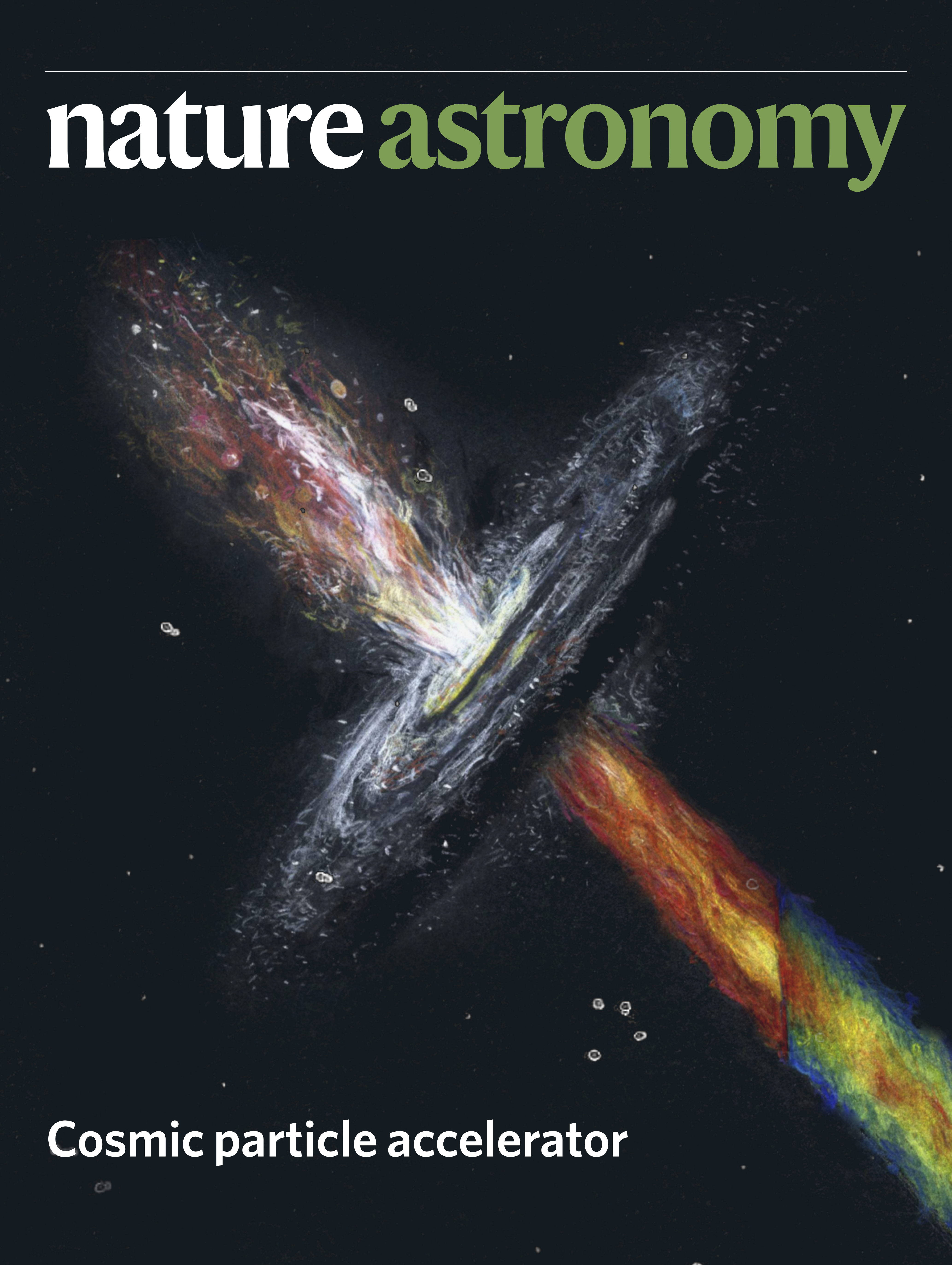} 
    \caption{Cover page of \textit{Nature Astronomy} Volume 4 Issue 2, February 2020. Image: E. Ampezzi, B. Del Piccolo, C. Schiavo (Liceo Artistico Modigliani, Padova) and E. Prandini (University of Padova). Cover Design: Allen Beattie.}
    \label{fig:cover}
\end{figure}

At the end of January 2020, the final cover became publicly available from the Nature website: \url{https://www.nature.com/natastron/volumes/4/issues/2}.  The cover is illustrated in Fig.~\ref{fig:cover}. 
This exceptional result, born as a simple school-university cooperation for a \texttt{scientific} conference, had a big resonance in the Italian media for a few weeks, until COVID-19 effects obscured our joy. Among others, we were interviewed by the national journals \texttt{Corriere della Sera} (see Fig.~\ref{fig:corriere}) and \texttt{Il Sole 24 Ore}.

\section{Closing remarks}
\noindent
Emily, one of the three students that designed the poster, said regarding this experience: 

\begin{quote}
\textit{
<<I did not expect our project to become the front page of an important scientific magazine such as Nature Astronomy, I was already very satisfied and proud by seeing our poster hanged on the walls at the conference venue. I am happy I took part of this experience because I found it both enriched the persons that took part in it, and brought together two far-away disciplines like astrophysics and arts>> [Emily]}
\end{quote}

\noindent As for the high school teachers that joined the project, these were their thoughts:

\begin{quote}
\textit{
<<This experience has been an unique occasion for students to experiment in an equipe. The sharing of knowledge and expertise has been increasingly refined, and brought to an art poster in which every details was precisely chosen on purpose.>> [D.~Mauri]}
\end{quote}

\noindent and 
\begin{quote}
\textit{
<<Every step of the project was stimulating, and I had to change often the point of view. While discussing with the astrophysicists, I felt like a student again, learning new things taught with passion about our Universe. The challenge was to traduce this passion into an art composition, which ought to be both good looking and scientifically accurate. When the poster was chosen by the Nature Astronomy journal, I admit I got moved!>> [B.~Montolli]}
\end{quote}

Finally, Dr Leila Zoia, of the Direction, Research and Outreach Sector of the Department of Physics and Astronomy in which the project was carried out, added:

\begin{quote}
\textit{<<This results was made possible by the strenuous work of the students, who strongly engaged during their stage, but also by the foresight of the University tutors that were able to put value to the students potential. Thanks to this close collaboration between different professionals, a powerful image was created, capable of intriguing and bringing the public closer to specialized concepts, fully hitting one of the objectives of scientific dissemination.>> [L. Zoia]}

\end{quote}

\bigskip
As for E.~Prandini, who followed the students closely during the project:

\begin{quote}
\textit{<<We \texttt{scientists} are always amazed by the targets of our \texttt{scientific} interest, and we often think of them through our knowledge and fantasies, or through our manuscripts, plots and graph. For once it has been amazing to look at those same targets with the eyes of \begin{test} artists!\end{test}>>}
\end{quote}

\section*{Acknowledgements}
 EP, MD, and MM acknowledge funding from Italian Ministry of Education, University and Research (MIUR) through the "Dipartimenti di eccellenza” project Science of the Universe. We thank Emily for her contribution to these proceedings.

\end{document}